\documentclass[aps,prx,twocolumn,superscriptaddress,floatfix]{revtex4}
\usepackage{graphicx}
\usepackage{dcolumn}
\usepackage{bm}
\usepackage{stmaryrd}
\usepackage{latexsym}
\usepackage{amssymb}
\usepackage{amsfonts}
\usepackage{amsmath}
\usepackage{fancybox}
\usepackage{color}
\usepackage{verbatim}
\usepackage{physics}
\usepackage{multirow}

\def\be{\begin{equation}}
\def\ee{\end{equation}}
\def\ber{\begin{eqnarray}}
\def\eer{\end{eqnarray}}

\begin{document}
%
%
\title{Gate tunable topological flat bands in twisted monolayer-bilayer graphene}
\author{Youngju Park}
\affiliation{Department of Physics, University of Seoul, Seoul 02504, Korea}
\author{Bheema Lingam Chittari}
\affiliation{Department of Physics, University of Seoul, Seoul 02504, Korea}
\author{Jeil Jung}
\email{jeiljung@uos.ac.kr}
\affiliation{Department of Physics, University of Seoul, Seoul 02504, Korea}

\begin{abstract}
We investigate the band structure of twisted monolayer-bilayer graphene (tMBG), 
or twisted graphene on bilayer graphene (tGBG), as a function of twist angles and perpendicular 
electric fields in search of optimum conditions for achieving isolated nearly flat bands.
Narrow bandwidths comparable or smaller than the effective Coulomb energies
satisfying $U_{\textrm{eff}} /W \gtrsim 1$ are expected for twist angles in the range of 
$0.3^{\circ} \sim 1.5^{\circ}$, more specifically in islands around 
$\theta \sim 0.5^{\circ}, \, 0.85^{\circ}, \,1.3^{\circ}$ for appropriate perpendicular electric field magnitudes and directions. 
The valley Chern numbers of the electron-hole asymmetric bands depend intrinsically on the details of the hopping 
terms in the bilayer graphene, and extrinsically on factors like electric fields or 
average staggered potentials in the  graphene layer aligned with the contacting hexagonal boron nitride substrate.
This tunability of the band isolation, bandwidth, and valley Chern numbers makes of tMBG a more 
versatile system than twisted bilayer graphene for finding nearly flat bands prone to strong correlations.
\end{abstract}
%


\maketitle
\section{Introduction} 
Van der Waals interfaces with small misalignment angles have emerged as a promising platform to 
study the correlated physics in layered materials where the low-energy bands are nearly flat due to extreme 
suppression of the Fermi velocities~\cite{deheer1,deheer2,deheer3,deheer4,deheer5,ugeda,ohta,lopes,lopes1,shallcross,shallcross1,
shallcross2,shallcross3,bistritzer,koshino,koshino1,jung,sanjose,sanjose1,sanjose2,Bistritzerprb,wangzf,schmidt, stephen2018,koshino2,vafek,vishwanath,vishwanath2}.  
Twisted bilayer graphene is the first representative system where the flatness of the bandwidth enhances the electron correlations 
leading to Mott-like insulating behaviors~\cite{mott1,mott2,mott3}
and signatures of superconductivity~\cite{super1,super2,super3}. 
A small twist angle of $\theta \approx 1^{\circ}$ between the van der Waals interfaces generates a long periodic moire superlattice on the order of $\sim10^{2}$ \r{A} giving rise to
a small mini moire Brillouin zone (MBZ) where the particles near the original Dirac points $K$ hybridize with each other through interlayer coupling~\cite{bistritzer,lopes,jung,dillonwong,abhay,nadj}. 
On the other hand, without any twist angle, moire superlattices are expected in graphene on hexagonal
boron nitride (G/BN) van der Waals interfaces due to a lattice mismatch of $\sim$~1.7$\%$~\cite{yankowitz}. 
The avoided gaps that appear at the Brillouin zone boundaries due to these moire potentials together with a
bandgap opening near the charge neutrality point~\cite{song,srivani2019} is an effective 
way to obtain isolated low-energy flat bands that have well defined valley Chern numbers. 
Systems where a primary Dirac point gap opens under the effect of an electric field has realizations in 
rhombohedral trilayer graphene on hexagonal boron nitride (TG/BN)~\cite{guorui1,guorui2,chittari,senthil} 
where the valley Chern number for either the valence or conduction bands $C_{v/c}$ are well-defined integers 
approximately proportional to layer number~\cite{chittari},
and twisted double bilayer graphene (tDBG)~\cite{tbbg, tbbg1, tbbg2, tbbg3, tbbg4, tbbg5, tbbg6} 
whose low energy bands assume valley Chern numbers range up to $\pm 4$ depending on the system parameters
and their bandwidths are reduced roughly by a factor of two
when compared to twisted bilayer graphene (tBG)~\cite{chittari2019, tbbg}.  
When the valley degeneracy is lifted and occupancy is polarized
~\cite{chittari, senthil,senthil2019,zaletel}
we can expect the onset of spontaneous quantum Hall phases even in the absence of magnetic 
fields~\cite{haldane1988, kanemele2005, nandkishore2010, jung2011, zhang2011} as verified in recent experiments~\cite{fengwang2019,mott3,Young_quantized}.

In this paper, we investigate the possibility of nearly flat bands in twisted monolayer graphene stacked on top of Bernal stacked bilayer graphene (tMBG), also known as twisted mono-bi graphene, 
where a linearly dispersing Dirac cone couples with a parabolic band.
This is the next simplest system to build experimentally after twisted bilayer graphene where only one twist angle interface
is present and 
can take advantage of the gate tunability of bilayer graphene where a gap can be opened by an external electric field.
The possibility of nearly flat bands in tMBG had been hinted in earlier theoretical works~\cite{senthil,ma2019topological,li2019electronic,levente} 
and the first series of experiments in tMBG have been reported very recently~\cite{shi2020tunable,chen2020electrically,polshyn2020}.
In this work we calculate various electronic structure properties relevant for interpreting experiments that had 
not been considered in earlier work,
including a complete phase diagram in the parameter space of twist angle and interlayer potential difference for the 
valley Chern number, the ratio $U_{\rm eff}/W$ between the Coulomb interaction versus bandwidth, and the impact of 
a finite bandgap in the monolayer graphene that can appear when it is aligned with a hexagonal boron nitride substrate layer.
This analysis is carried out using an improved full bands continuum Hamiltonian model that incorporates the remote hopping terms and the interlayer coupling matrix elements accounting for out of plane relaxation effects.
 
%
\begin{figure}[thb]
\includegraphics[width=8.5cm,angle=0]{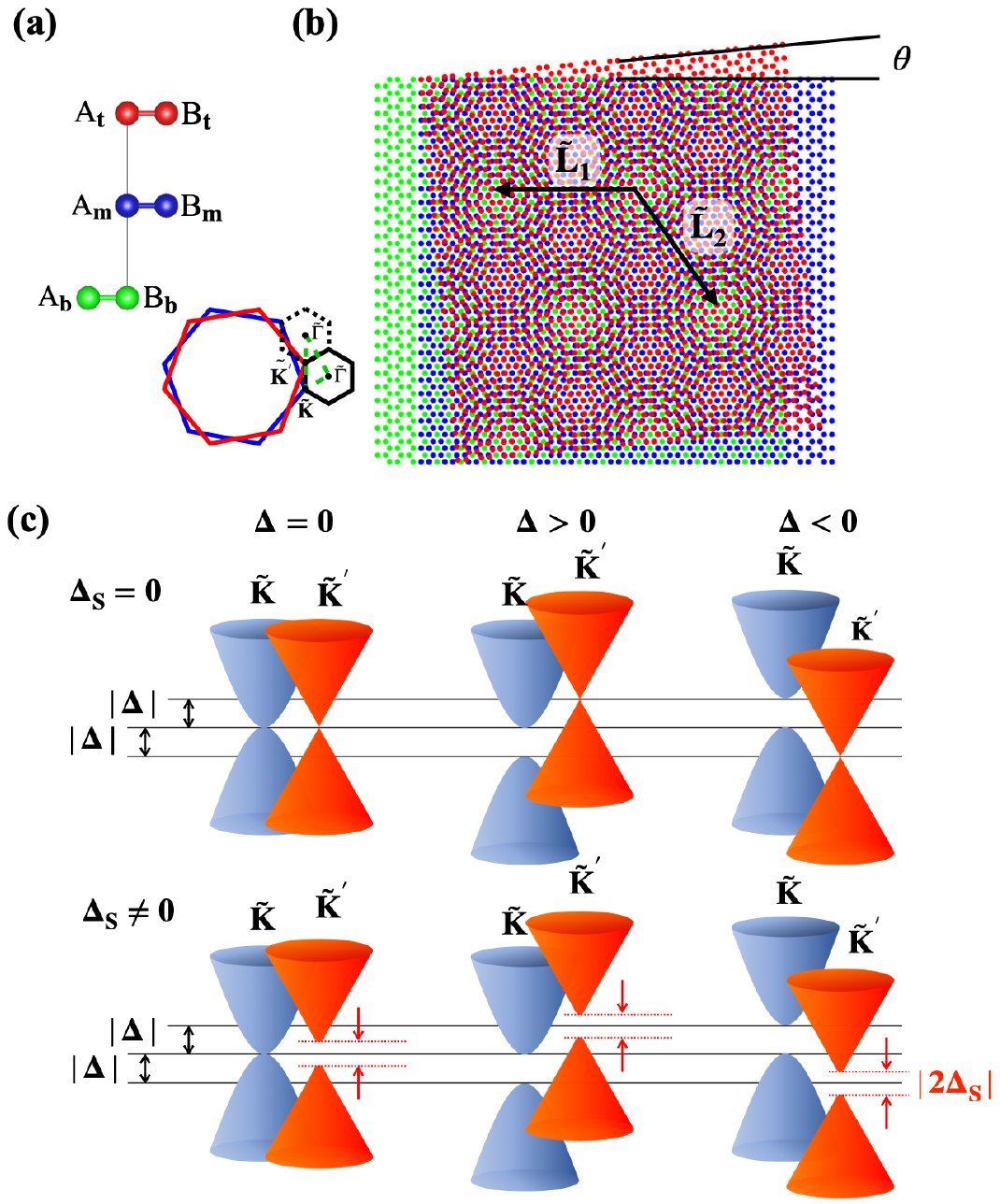} 
\caption{(color online) 
Schematic diagram of twisted monolayer graphene stacked on top of a Bernal stacked bilayer graphene (tMBG). 
(a) The side-view and,  (b) top-view of the tMBG atomic structure. The $A_l$ and $B_l$ are two sublattices 
of {\em $l^{th}$} layer, {\em l = }b (bottom), m (middle), t (top). 
A relative twist angle $\theta$ between the monolayer and bilayer graphene gives rise to a moire pattern 
lattice vectors $\tilde{L}_i$. The mini moire Brillouin Zone (MBZ) is illustrated by a black hexagon, 
along with the MBZ high symmetry points. The red and blue hexagons correspond to the first Brillouin zone 
of monolayer and bilayer graphene. (c) We schematically illustrated the effect of an applied electric field
by means of the interlayer potentials $\pm \Delta$ introduced between the bottom and top graphene layers,
and the staggered potential $\Delta_S$ in the top monolayer graphene due to alignment 
with hexagonal boron nitride giving rise to $\Delta_S < 0$ for BN alignment 
(for graphene's $A_t$ and $B_t$ sublattices aligned with boron and nitrogen sites) and to $\Delta_S > 0$ for NB alignments.
}
\label{Fig:schematic}
\end{figure}
%

The manuscript is structured as follows. In Sec.~\ref{hamiltonian} we define the model Hamiltonian of the system. 
The Sec.~\ref{results} is devoted to the discussion and presentation of various results in the parameter space of twist 
angle $\theta$ and interlayer potential difference proportional to $\Delta$ for a number of observables of the system including the bandwidth $W$, 
the $U_{\textrm{eff}}/W$ ratio between the effective Coulomb interaction $U_{\textrm{eff}}$ versus bandwidths $W$,
the (local) density of states,
the valley Chern numbers of the low lying energy bands and associated Berry curvatures.
In Sec.~\ref{conclusions}, we close the manuscript with the summary and conclusions.

\section{Model Hamiltonian} 
\label{hamiltonian}
Our model Hamiltonian is based on the moire bands theory~\cite{bistritzer,jung}
that expands beyond the proposal of Bistritzer-MacDonald of twisted bilayer graphene
by incorporating information from first-principles calculations for both intralayer and interlayer moire patterns~\cite{jung}. 
The description of our twisted monolayer-bilayer (tMBG) system is improved with respect to 
earlier works in Ref.~\cite{senthil,ma2019topological,li2019electronic} by considering full bands continuum Hamitonian
that incorporates renormalized out of plane tunneling parameters compatible with exact exchange and random 
phase approximation (EXX+RPA) equilibrium distances for different local stacking~\cite{tbbg} and by incorporating 
the remote hopping parameters in the bilayer graphene~\cite{accuratebilayer}. 
The Hamiltonian model of tMBG with a twist angle $\theta$ is given by
   \begin{equation}
   \begin{aligned}
   H_{\rm tMBG}(\theta) = 
    \begin{pmatrix} h_{b}^{-} + \Delta_{b}  & t^{-} & 0   \\ 
    t^{- \dagger} & h_{m}^{-}  + \Delta_{m}&  T(\bm{r}) \\
    0 & T^{\dagger}(\bm{r}) & h_{t}^{+} + \Delta_{t} \end{pmatrix}, 
   \end{aligned}
\label{Eq:Hamiltonian}
   \end{equation}
where the Hamiltonian elements for the bilayer graphene (BG) consisting of the bottom and middle layers have a a small onsite
potential at the higher-energy dimer sites proportional to $\delta \sim 15$~meV~\cite{accuratebilayer}
   \begin{equation}
   \begin{aligned}
   h_{b}^{-} &= h(- \theta / 2) + \delta(\sigma_z \, + \, \mathbb{I})/2, \\   
   h_{m}^{-} &= h(- \theta / 2) + \delta(\sigma_z \, - \, \mathbb{I})/2, \\
   h_{t}^{+} &= h( \theta / 2) 
   \end{aligned}
   \end{equation}
where $h_{l}^{\pm} = h(\pm \theta / 2)$ uses the graphene Hamiltonian 
with a rotated phase $ h(\theta/2)= \upsilon_{\rm F} \hat{R}_{-\theta/2} \bm{p} \cdot \sigma_{xy}$ using 
graphene sublattice pseudospin Pauli matrices $\sigma_{xy} = (\sigma_x,\sigma_y)$ and $\sigma_z$ and $\mathbb{I}$ is a $2\times2$ identity matrix. The rotation $\hat{R}_{-\theta/2}$ of the $\bm{p}=(p_x,p_y)$ vector which is the momentum measured from $K$ valley introduces a phase
$e^{i \theta_{\bm p}} \rightarrow e^{i (\theta_{\bm p} - \theta/2)} $ proportional to $\theta/2$.
The valley index $\nu=\pm 1$ denote one of the two different Dirac points $\bm{K}_{\nu} = (\nu 4 \pi/3 a, 0)$ of the unrotated monolayer/bilayer graphene BZ. The result sin this work will be for the $\nu = 1$ 
macro-valley $K$ unless stated otherwise. 
The Fermi velocity $\upsilon_{\rm F} = \upsilon_0$ we used is defined through $v_i = \sqrt{3}|t_i|a/2 \hbar$ and 
relates $\upsilon_{\rm F}$ to the intralayer nearest-neighbor 
hopping parameter value $t_0=-3.1$~eV 
that is enhanced by $\sim$20\% over the $t_0 = -2.6$~eV obtained within the local density approximation (LDA)~\cite{accuratebilayer} is a commonly used value in the graphene literature 
that is appropriate for band models that do 
not explicitly include non-local Coulomb interaction terms that can explicitly 
enhance the Fermi velocity. 
This enhanced value allows to capture the experimental moire bands features when the interlayer 
tunneling strengths are tuned to the first principles DFT calculation values for proper interlayer separation distances~\cite{tbbg,dillonwong}.
The interlayer coupling terms within BG are given by 
 \begin{equation}
   \begin{aligned}
   t^{\pm} &= 
    \begin{pmatrix} - \upsilon_4 \pi^{\pm} & - \upsilon_3 \pi^{\pm \dagger}  \\ 
                    t_1 & - \upsilon_4 \pi^{\pm}\end{pmatrix},
   \end{aligned}
   \end{equation}
where $\pi^{\pm} =  p e^{ i(\theta_{\bm p} \mp \theta/2)}$ 
contains the rotation dependent phase factor. 
The interlayer coupling matrix $t^{\pm}$ within the BG contains the perpendicular coupling term $t_1=~0.36$~eV 
and the remote hopping contributions, including $t_3=0.283$~eV and $t_4=0.138$~eV adopted from the accurate tight-binding model of the bilayer graphene and cause trigonal warping and the electron-hole asymmetry~\cite{accuratebilayer}. In the minimal model that we also present for comparison purposes, 
we ignore those remote hopping terms and the site-potential difference $\delta$ between the dimer and non-dimer sites in the BG Hamiltonian.

The interlayer coupling terms between the twisted G and BG systems are given by
   \begin{equation}
   \begin{aligned}
   T(\bm{r}) = \sum_{j=0,\pm} \: e^{-i\, \bm{Q}_j \cdot \bm{r}} \: T^{j},
   \end{aligned}
   \end{equation}
   where the three $\bm{Q}_j$ vectors are $\bm{Q}_0 = K \theta (0,-1)$ and $\bm{Q}_{\pm} = K \theta (\pm \sqrt{3}/2,1/2)$ with $K = 4 \pi/3a$, and $T^{j}$ are
   \begin{equation}
   \begin{aligned}
   T^{0} = \begin{pmatrix} \omega^{\prime}  & \omega \\ \omega  & \omega^{\prime}\end{pmatrix}, \:\:
   T^{\pm} = \begin{pmatrix} \omega^{\prime}  & \omega\, e^{\mp i 2\pi \nu/3} \\ \omega\, e^{\pm i 2\pi \nu/3}  & \omega^{\prime}\end{pmatrix}, 
   \end{aligned}
\label{Eq:Tcoupling}
   \end{equation}
   where $\omega = \omega_{BA^{\prime}} = t_1 /3$, and $\omega^{\prime} =\omega_{AA^{\prime}}= (-0.1835 {t_1}^2 + 1.036 {t_1} - 0.06736)/3$. 
   We chose the different values for $\omega$ and $\omega^{\prime}$ to consider the effect of out of plane atomic relaxation. 
   Those parameters are adopted from EXX+RPA fitting values in the supplementary materials of Ref.~{\cite{tbbg}. 
   We use the interlayer potential values $\Delta$ ($\Delta_b$ for the bottom, $\Delta_t$ for top and $\Delta_m$ for the middle layer) as 
\begin{equation}
   \begin{aligned}
   \Delta_{b} &= -\Delta \mathbb{I}, \\   
   \Delta_{m} &= \mathbb{O} , \\
   \Delta_{t} &=  \Delta \mathbb{I} + {\Delta_S} \, \sigma_z.
   \end{aligned}
\label{Eq:potential}
   \end{equation}
Following the schematic representation in Fig.~\ref{Fig:schematic}, in loosely coupled sufficiently large twist angle tMBG
it is possible to identify respectively simultaneous traces of quadratic bands from BG and linear bands from G 
near $\tilde{K}$ and $\tilde{K}^{'}$ of the moire Brillouin zone (MBZ). 
Their degree of hybridization changes with the interlayer coupling term $\omega$ and twist angle magnitude.
The presence of $\Delta_S$ contributes to opening the bandgap at the $\tilde{K}^{\prime}$ of the monolayer graphene and, although weakly, 
facilitates the isolation of the low energy nearly flat bands. 

\begin{figure}[!htb]
\includegraphics[width=8.5cm,angle=0]{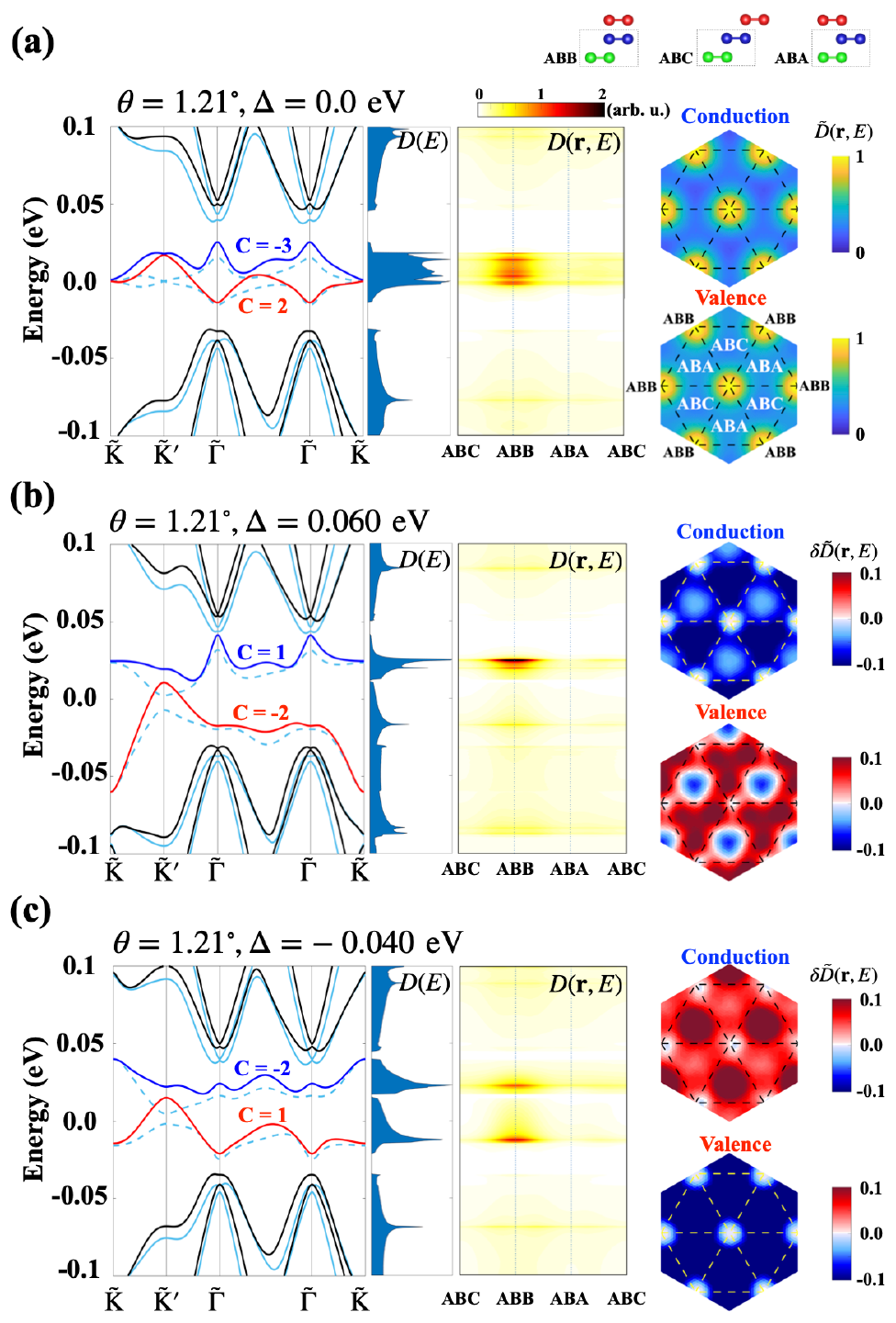} 
\caption{(color online) 
Band structures for different values of interlayer potentials (a) $\Delta = 0$ eV, (b) $\Delta = 0.060$ eV and (c) $\Delta = -0.040$ eV
for a twist angle $\theta=1.21^{\circ}$ obtained using minimal/complete models 
illustrated by the dotted/solid lines together with the complete model's 
density of states $D(E)$, local density of states $D({\mathbf r},E)$
for different local commensurate stacking regions ABB, ABC, and ABA,
and their differences
$\delta \tilde{D}({\mathbf r},E) = \tilde{D}_{\Delta}({\mathbf r},E) - \tilde{D}_{0}({\mathbf r},E)$ for finite 
$\Delta$ interlayer potentials
where
$\tilde{D}({\mathbf r}, E) = D({\mathbf r},E) / {\rm max}(| D({\mathbf r},E)| )$ are the normalized LDOS evaluated at 
the van Hove singularity energies $E = E_{\rm VHS}$ of the nearly flat bands.
The isolated low-energy bands of valence/conduction bands are represented by red/blue solid lines 
accompanied by their valley Chern numbers. 
On the right we represent the real space plot of normalized LDOS at 
the van Hove singularities (VHS) of conduction(blue)/valence(red) for $\Delta = 0$~eV,
and their differences for $\Delta = 0.060$~eV and $\Delta = -0.040$~eV
showing opposing contrasts for the conduction and valence bands depending on the sign of $\Delta$.
}
\label{Fig:Bands1}
\end{figure}

\section{Results and discussions}
\label{results}

\subsection{Flat bands} 
Here we explore the parameter space of twist angles and electric field induced interlayer potentials 
$\Delta$ in search of the optimum conditions
for the generation of isolated flat bands where we can expect to see correlation effects. 
Inspection of minimal model moire bands in tMBG shows that narrow bandwidths
of the order of $\sim$1~meV are achievable near the magic angle $\theta = 1.07^{\circ}$ comparable to
the bandwidths seen for the flat bands in the minimal model tBG~\cite{chittari2019, tbbg}, tDBG~\cite{tbbg} 
and twisted multilayer graphene~\cite{tmg_DaiXi}.
When the remote hopping terms are added in the BG the bandwidths of the flat bands at the twist angle 
$\theta=1.07^{\circ}$ broaden up to 15$\sim$20~meV. 
While the bandwidths see an increasing trend with twist angles,
introducing interlayer potential differences of magnitudes $| \Delta | \lesssim 50$~meV 
often allows to reduce their bandwidths and allows to isolate them by opening a primary and secondary gaps,
see appendix Fig.~\ref{Fig:BW-supple1} and \ref{Fig:BW_GAP-supple} for the relevant illustrations.
Hence, even if the broadened bandwidths are less favorable than that of magic angle 
tBG due to the remote hopping terms, the presence of a finite interlayer potential difference $\Delta$
allows to further tune the low energy bandwidths in tMBG in a manner similar to tDBG~\cite{tbbg} or 
massive tBG~\cite{srivani2019}
when a bandgap opens at the primary Dirac point.
While the conduction bandwidths tend to be narrower than those of the valence bands,
we notice a strong electric field direction dependent asymmetry of the bandwidths when compared to tBG and tDBG
that leads to minimum conduction bandwidths for $\Delta > 0$ and minimum valence bandwidths for $\Delta<0$. 
This trend can be verified to remain valid up to twist angles as large as $\sim1.5^{\circ}$  beyond which the bandwidths increase steeply above $\sim$50~meV, see appendix Fig.~\ref{Fig:BW_GAP-supple}.

The isolation of the flat bands due to the electric fields allows to characterize the valley resolved 
Chern number $C_n$ of the $n$-th band through
 \begin{equation}
   \begin{aligned}
   C_n &= \int_{\rm MBZ} d^2 \bm{{k}} \quad \Omega_n(\bm{{k}})/(2 \pi),
   \end{aligned}
   \label{Eq:Chern}
   \end{equation}
where the Berry curvature $\Omega_n(\bm{{k}})$ is defined through~\cite{berry_rmp}
   \begin{equation}
   \begin{aligned}
   \Omega_n(\bm{{k}}) &= -2 \sum_{n^{\prime} \neq n }\mathrm{Im} \left[\frac{\mel{u_n}{\frac{\partial H}{\partial {k}_x}}{u_{n^{\prime}}}\, \mel{u_{n^{\prime}}}{\frac{\partial H}{\partial {k}_y}}{u_n}}{(E_{n^{\prime}} - E_n)^2}\right],
   \end{aligned}
   \label{Eq:Berry}
   \end{equation}
where $\ket{u_n}$ is the moire Bloch states, and $E_n$ are the band energies.
For instance, the low-energy valence band of tMBG at $\theta = 1.21^{\circ}$ has a valley Chern number $C = -3$, 
and $C = 2$ for the conduction band in the absence of an electric field ($\Delta = 0$).
Upon application of an electric field $\pm \Delta$, we can trigger changes in the topological numbers 
both for conduction and valence bands.
See Fig.~\ref{Fig:Bands1} and the appendix Fig.~\ref{Fig:Bands-supple} for a closer illustration of the 
nearly flat bands prone to strong correlations for select 
twist angles 0.51$^{\circ}$, 0.85$^{\circ}$, 1.21$^{\circ}$, 1.31$^{\circ}$,  and 1.41$^{\circ}$. 
In Fig.~\ref{Fig:Bands1}, we illustrated the band structures at $\theta=1.21^{\circ}$ for 
three different electric fields $\Delta = 0.0, 0.060, -0.040$~eV where the bandwidth of 
valence or conduction bands reach minima values. 
Switching the direction of the interlayer potential differences $\Delta$ 
allows to achieve different valley Chern numbers $-2/1$ for the valence/conduction bands, respectively, and whose values can be reversed with the sign of the electric field. 

\begin{figure*}[!htb]
\includegraphics[width=15.5cm,angle=0]{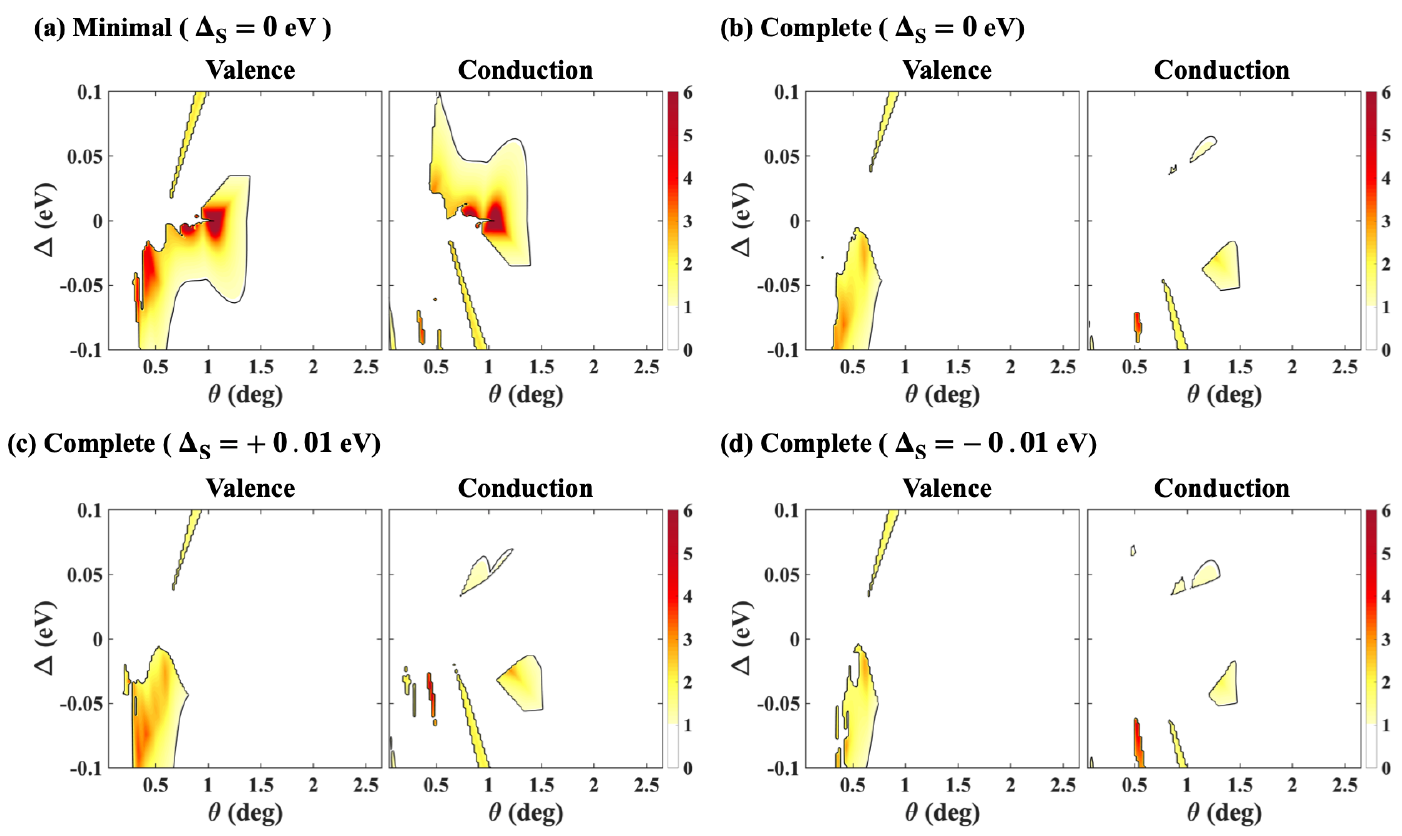} 
\caption{(color online) The phase diagram of ratio between the effective screened Coulomb potential $U_{\mathrm{eff}}$ and bandwidth $W$ ($U_{\mathrm{eff}}/W$) as a function of $\theta$ and $\Delta$ in tMBG, where $U_{\mathrm{eff}}/W>1$ indicates the Coulomb-interaction driven ordered phases. (a) and (b) compare the $U_{\mathrm{eff}}/W$ ratio obtained using minimal and complete model. The unequal colormap of, $U_{\mathrm{eff}}/W$ between the valence and conduction band indicates the strong particle-hole asymmetry in tMBG. The complete model reduces the area of $U_{\mathrm{eff}}/W>1$ when compared to the minimal model. The colormap distribution of $U_{\mathrm{eff}}/W$ is changed slightly with the staggered potential $\Delta_S > 0$, (c) $\Delta_S = 0.01$ eV and (d) $\Delta_S = - 0.01$ eV. The area of $U_{\mathrm{eff}}/W>1$ is enlarged near twist angle of $\theta\approx0.5^{\circ}$ and $\theta\approx1.2^{\circ}$ for $\Delta_S>0$ when compared to $\Delta_S = 0$ case within the complete remote hopping parameters model, meanwhile the area $U_{\mathrm{eff}}/W>1$ is reduced when $\Delta_S < 0$. }
\label{Fig:Ueff_W}
\end{figure*}

\begin{figure}[!htb]
\includegraphics[width=8.5cm,angle=0]{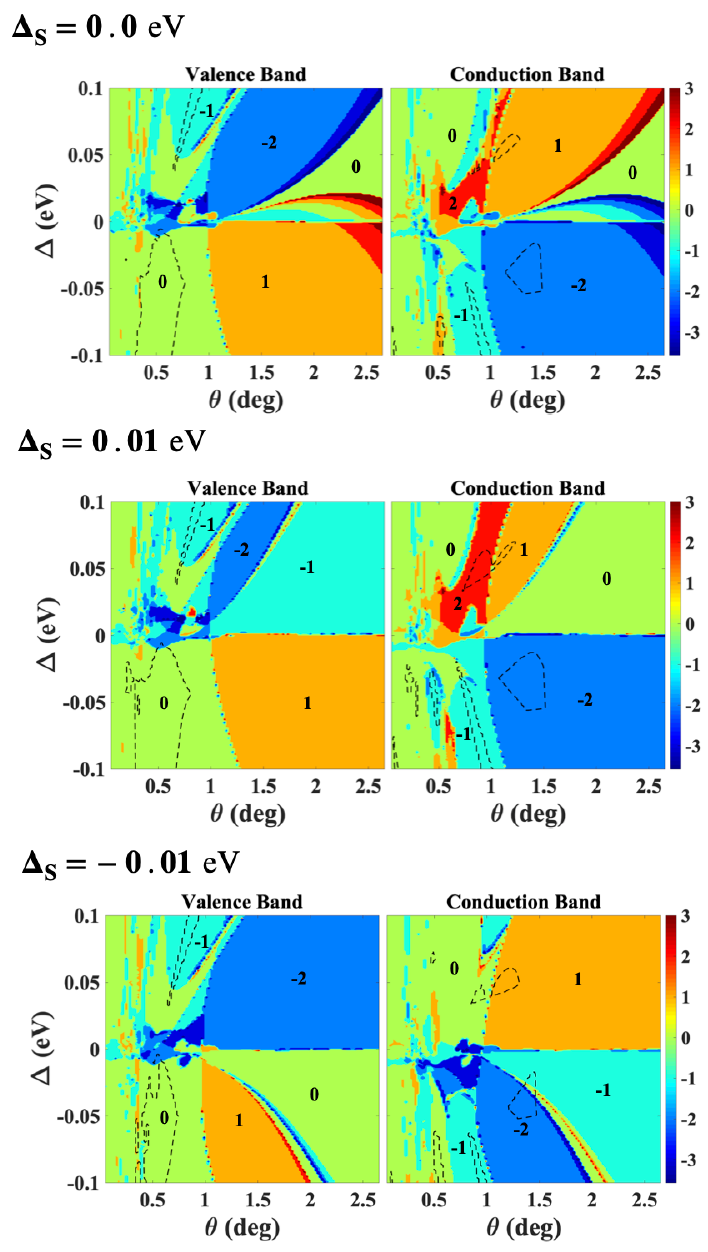} 
\caption{(color online) 
Phase diagram of the valley Chern numbers in tMBG for the low-energy conduction and valence bands 
as a function of twist angle $\theta$ and interlayer potentials $\Delta$ for three values of sublattice 
staggering potential $\Delta_S$ in the monolayer. 
The different valley Chern numbers ($C = 0, \pm 1, \pm 2, \pm 3$) 
for the valence/conduction bands as a function applied electric field $\Delta$
indicates the possibility of achieving topological phase transitions for various system parameters. 
The large $C = \pm 3$ phases disappear for $\Delta_S \neq 0$. 
The black dashed lines encloses the respective strong correlation regions satisfying $U_{\mathrm{eff}}/W>1$
in Fig.~\ref{Fig:Ueff_W}.}
\label{Fig:Chern}
\end{figure}

Isolation of the low energy flat bands from the neighboring bands can be aided through $\Delta$ and $\Delta_S$ parameters that can open gaps
in the bilayer and monolayer components of tMBG.
As we mentioned earlier, when the monolayer and bilayer bands are loosely coupled for large twist angles,
we can distinguish the linearly dispersing bands at $\tilde{K^{\prime}}$ of the monolayer graphene
and the quadratic bands at $\tilde{K}$ for the bottom BG, see Eq.~(\ref{Eq:potential}). 
While gaps in the bilayer component near $\tilde{K}$ can be opened with $\Delta$, 
it won't open for the Dirac cone of monolayer graphene at $\tilde{K}^{\prime}$ which instead can be achieved 
by aligning an hBN layer~\cite{massivedf,massivegraphene,origingap}.
This effect can be captured through a site-potential difference $\Delta_S$ representing an average constant 
staggered potential difference between the two sub-lattices $A_t$ and $B_t$ 
and can be used as an additional control parameter for tMBG. 
In the appendix Fig.~\ref{Fig:BW_GAP-supple} (c) and (d), we illustrate how the bandwidths, primary and secondary band gaps 
change for different signs of the sublattice staggered potentials $\Delta_S >0 $ or $\Delta < 0$ 
depending on the relative alignment of hBN contacting the monolayer graphene.
We notice that a staggered potential $\Delta_S$ generally improves the isolation of the bands
even if the total bandwidths are similar to the  $\Delta_S=0$ case.
The staggered potential due to hBN has also an effect on the topology of the low energy bands. 
In Fig.~\ref{Fig:Bands2}, we presented the band structures of tMBG at $\theta=1.07^{\circ}$ and 
$\theta=1.21^{\circ}$ with staggered potentials of $\Delta_S = \pm 0.01$~eV for select values of electric fields. 
Here, we observe that the sign differences in $\Delta_S$ introduces small changes in the band structure but leads to non-negligible changes in the local density of states and, at times, to the valley Chern numbers 
which in turn can impact the transport properties. 

\subsection{Local density of states}\label{Sec:LDOS}
The local density of states (LDOS) maps in tMBG behave analogously to those of tBG in the fact that the charge densities 
tend to concentrate at local stacking configurations where the monolayer and top bottom layer units cells are on top of each other at the ABB stacking configurations, see Fig.~\ref{Fig:Bands1} (a). 
The sensitivity of the flat bands to interlayer potential differences $\Delta$ and sublattice staggering potential $\Delta_S$ discussed earlier
suggests that the electron localization properties can also be tuned by means of those system control knobs. 
For this purpose we define the normalized LDOS difference as
$\delta \tilde{D}({\mathbf r},E) = ( \tilde{D}_{\Delta_{(S)}}({\mathbf r},E) - \tilde{D}_{0}({\mathbf r},E) )$
for finite interlayer potential $\Delta$ or staggered potential $\Delta_{S}$
where the tildes indicate normalization $\tilde{D}({\mathbf r},E) = D({\mathbf r},E) / {\rm max}( D({\mathbf r},E) )$
and the energy $E$ is chosen to sit at the van Hove singularity of the flat band under consideration.
The increase and decrease of LDOS happen mainly at the ABC and ABA stacking locations for variable $\Delta$
as noted in Fig.~\ref{Fig:Bands1}.
For $\Delta > 0$ that increases the population of the electrons at the bottom layer we observe 
an overall increase of valence band electrons at  ABA and depletion at ABC, and general depletion at both ABA and ABC 
for the conduction band electrons. 
Reversing the electric field for $\Delta < 0$ favoring the accumulation of electrons at the top layer
reverses this overall behavior but enhancing the population at both ABA and ABC stacking locations 
for the conduction bands and depleting for the valence bands. 
Significant changes in the LDOS maps are seen also when we introduce a finite $\Delta_S$ as illustrated in 
Fig.~\ref{Fig:Bands2} where the height of the LDOS at ABB stacking can change 
significantly depending also on the other system parameters. 
This observation is in keeping with the fact that the electron wave functions of the flat bands locate primarily at the low energy carbon sites of the bilayer and at the carbon sublattice of the monolayer right on top of the bilayer low energy site, see appendix Fig.~\ref{Fig:LDOS-supple}.


\subsection{Strong correlations at large U$_{\textrm{eff}}$/W regions}\label{Sec:Ueff}
The $U_{\textrm{eff}}/W$ diagram in Fig.~\ref{Fig:Ueff_W} relating the Coulomb interaction strength $U_{\textrm{eff}}$
against the bandwidth $W$ show that maxima spots are possible for twist angles
$\theta\approx 0.3\sim0.8^{\circ}$ for the valence flat bands and $\theta\approx 1.1\sim 1.5^{\circ}$ for the 
conduction bands under appropriate interlayer potentials $\Delta$, generally more favorable for 
$\Delta < 0$ when the electric fields favor accumulation of electrons in the top layer,
although either field directions can generate isolated flat bands for the larger twist angles.
Further introduction of the staggered potential on the monolayer 
of $\Delta_S$ modifies the $U_{\textrm{eff}}/W >1$ phase diagram generally enlarging and reducing 
its area respectively for positive and negative $\Delta_S$, in particular near 
$\theta\approx0.5^{\circ}$ and $\theta\approx1.3^{\circ}$.
In order to estimate the Coulomb interaction strength we adopted the formula for the effective 
three-dimensional screened Coulomb potential which can be written as 
   \begin{equation}
   \begin{aligned}
   U_{\textrm{eff}} = \frac{e^{2}}{4 \pi \epsilon_{r} \epsilon_{0}  \tilde{L}} \textrm{exp}(- \tilde{L} / \lambda_{D}),
   \end{aligned}
\label{Eq:Ueff}
   \end{equation}
where we used $\epsilon_{r}=4$ and the Debye length $\lambda_{D} = 2 \epsilon_{0} /e^{2} D(\delta_{p}, \delta_{s})$ which includes the 2D density of states $D(\delta_{p}, \delta_{s}) = 4 [|\delta_{p}|\, u(-\delta_{p}) \,+\, |\delta_{s}|\, u(-\delta_{s})]/(W^{2} A_{M})$. 
The moire length $\tilde{L} \sim a/\theta$ depends on graphene's lattice constant and 
twist angle $\theta$, $A_{M} = \sqrt{3}\tilde{L}^2 / 2$ is the moire supercell area, $W$ denotes the band widths, and $u(x)$ is the heaviside step function such that $u(-\delta_{p/s})$ enhances screening 
in the presence of band overlap for negative values of the primary and secondary gaps $\delta_{p/s}$.
From the formula in Eq.~(\ref{Eq:Ueff}), we can identify the parameter space in $\theta$ and $\pm \Delta$ with large $U_{\textrm{eff}}/W > 1$ prone to strong correlations for both valence and conduction bands. 
In Fig.~\ref{Fig:Ueff_W} we compare the results of $U_{\textrm{eff}}/W$
between the minimal model and complete model, and consider the staggered potential ($\Delta_S = \pm 0.01$~eV) 
that can be introduced by alignment of the monolayer graphene with hBN. 
The phase diagram for $U_{\textrm{eff}}/W >1$ has a strong electron-hole asymmetry and sign 
dependence to $\Delta$ that is naturally expected from the structural asymmetry of tMBG
and was not observed in tBG~\cite{chittari2019} and tDBG~\cite{tbbg}.
We notice that even the symmetry  of the $U_{\textrm{eff}}/W$ diagram
between the electron and hole flat bands for opposite $\Delta$ signs 
present for the minimal model is destroyed when the remote hopping terms in the BG Hamiltonian
introduce significant overlap between neighboring bands. 

\subsection{Topological moire bands} \label{Sec:Chern}
Well defined non-trivial valley Chern numbers are expected in isolated moire bands and they are
believed to underlie the spontaneous quantum Hall effects observed in experiments when the degeneracy 
of the flat bands are lifted by Coulomb interactions~\cite{fengwang2019,mott3,Young_quantized}. 
The valley Chern numbers that we calculated following Eq.~(\ref{Eq:Chern}) range the values of
$C_{v/c} = 0, \pm 1, \pm 2, \pm3$ for the low-energy valence and conduction bands
in the parameter space of $\theta$ and $\Delta$ as we show in the Fig.~\ref{Fig:Chern} 
and is further modified through $\Delta_S$ as shown in appendix Figs.~\ref{Fig:Bands-supple1}-\ref{Fig:BandsChern}. 
For the minimal model (not shown) a certain degree of symmetry is preserved in the Chern number 
phase diagram for the valence and conductions bands of opposite $\Delta$ signs 
as we had noted for the $U_{\mathrm{eff}}/W$ phase diagram. 
With the remote hopping terms included the valley Chern 
numbers for the most promising flat bands take values of $C = -2$ for the conduction bands
in the vicinity of  $\sim 1.3^{\circ}$ for  
$\Delta<0$ and $C = 1$ for $\Delta > 0$. The Chern number is $C = -1$ 
for the flat bands expected at $\theta \sim 0.85^{\circ}$ for valence and conduction bands with opposite electric fields,
and a Chern number of $C=0$ is expected for the valence flat bands near $\theta \sim 0.5^{\circ}$ for $\Delta <0$.
We observe contrasting behaviors for $\Delta >0 $ with the negative and positive
valley Chern numbers of $C = \mp1, \mp2, \mp3$ respectively for most valence and conduction bands,
while for $\Delta<0$ they assume values of $C=0,\pm1,\pm2$. 
The strong electron-hole asymmetry  together with the large tunability of the band structures 
with the electric fields makes of tMBG an intersting system where it is possible to 
access multiple valley Chern number regions.

\section{Summary and conclusions}
\label{conclusions}
In this paper we investigated the conditions for the onset of strongly correlated topological flat bands 
in twisted monolayer graphene on Bernal stacked bilayer graphene (tMBG)
as a function of twist angle $\theta$ and interlayer potential $\Delta$ parameters.
Proper inclusion of the remote hopping terms in the bilayer graphene enhances broadening of the low energy
bandwidths and introduces overlap between neighboring bands reducing the parameter space 
of strongly correlations where the ratio $U_{\mathrm{eff}}/W>1$ between 
the effective Coulomb interactions and bandwidths remain large. 
However, a finite interlayer potential $\Delta$ by a perpendicular electric field
can reduce the bandwidth $W$ and 
isolate the low energy bands by opening primary and secondary band gaps. 
The system responds asymmetrically to the electric field direction
for both valence and conduction bands
resulting in multiple islands of large $U_{\rm eff}/W \gtrsim 1$ regions.
We have  summarized in Fig.~\ref{Fig:Ueff_W} together with the valley Chern numbers in Fig.~\ref{Fig:Chern}  
the conditions for strong correlations in the valence or conduction bands 
that we expect for twist angles in the range of $0.3^{\circ} \sim 0.8^{\circ}$ 
and $1.1^{\circ} \sim 1.5^{\circ}$ for appropriate interlayer potential differences, more specifically in islands around 
twist angles of $\theta \sim 0.5^{\circ}, \, 0.85^{\circ}, \,1.3^{\circ}$.
The bandwidths are narrowest for small twist angles and
they quickly increase above $W \gtrsim$50~meV for twist angles larger than $\theta \gtrsim 1.5^{\circ}$
precluding strong correlations for large twist angles despite that the bands can be isolated through 
a perpendicular electric field.
This phase diagram can be further altered by an average sublattice staggering potential $\Delta_S$ 
that can be introduced in the monolayer graphene through alignment with a hexagonal boron nitride layer.
We have shown that the tunability of the bandwidths, band isolation, 
and valley Chern numbers through the twist angle and interlayer potential differences
makes of tMBG a more versatile system than twisted bilayer graphene 
for generating nearly flat moire bands prone to strong correlations.

\section{Acknowledgments.}
This work was supported by the Samsung Science and Technology Foundation under project 
No. SSTF-BA1802-06 for Y. P. and the Basic Science Research Program through the National 
Research Foundation of Korea (NRF) funded by the Ministry of Education Grants No. 
2018R1A6A1A06024977 and Grant No. NRF-2020R1A2C3009142 for B.L.C.,
and by the Basic Study and Interdisciplinary R\&D Foundation Fund of the University 
of Seoul (2019) for J.J. 

\bibliography{myref.bib}



\begin{appendix}
\renewcommand\thefigure{A\arabic{figure}}    
\setcounter{figure}{0}    
\section*{Appendix}
\section*{A1. Bandwidth as a function of twist angle $\theta$}
\label{appendix_bandwidth}
The bandwidth of the $n$-th band at a given twist angle is calculated as $W = E^n_{\rm max} - E^n_{\rm min}$.  We show the bandwidth variation as a function of twist angle $\theta$ for tMBG in Fig~\ref{Fig:BW-supple1}. As it is observed in the twisted bilayer graphene (tGB), the bandwidth of the low energy bands decreases together with twist angle and reaches local minima at certain angles. In the minimal model of tMBG the bandwidth of the low-energy bands (valence/conduction) decreases with twist angle reaching the lowest bandwidth at $\theta=1.07^{\circ}$.
In contrast to the minimal model, the complete model that includes the remote hopping terms shows 
broadened bandwidths
that have markedly different behaviors depending on the two signs (negative/positive) of $\Delta$, indicating the electric field direction dependent asymmetry on the bandwidth in tMBG.  

\begin{figure*}[!htb]
\includegraphics[width=16cm,angle=0]{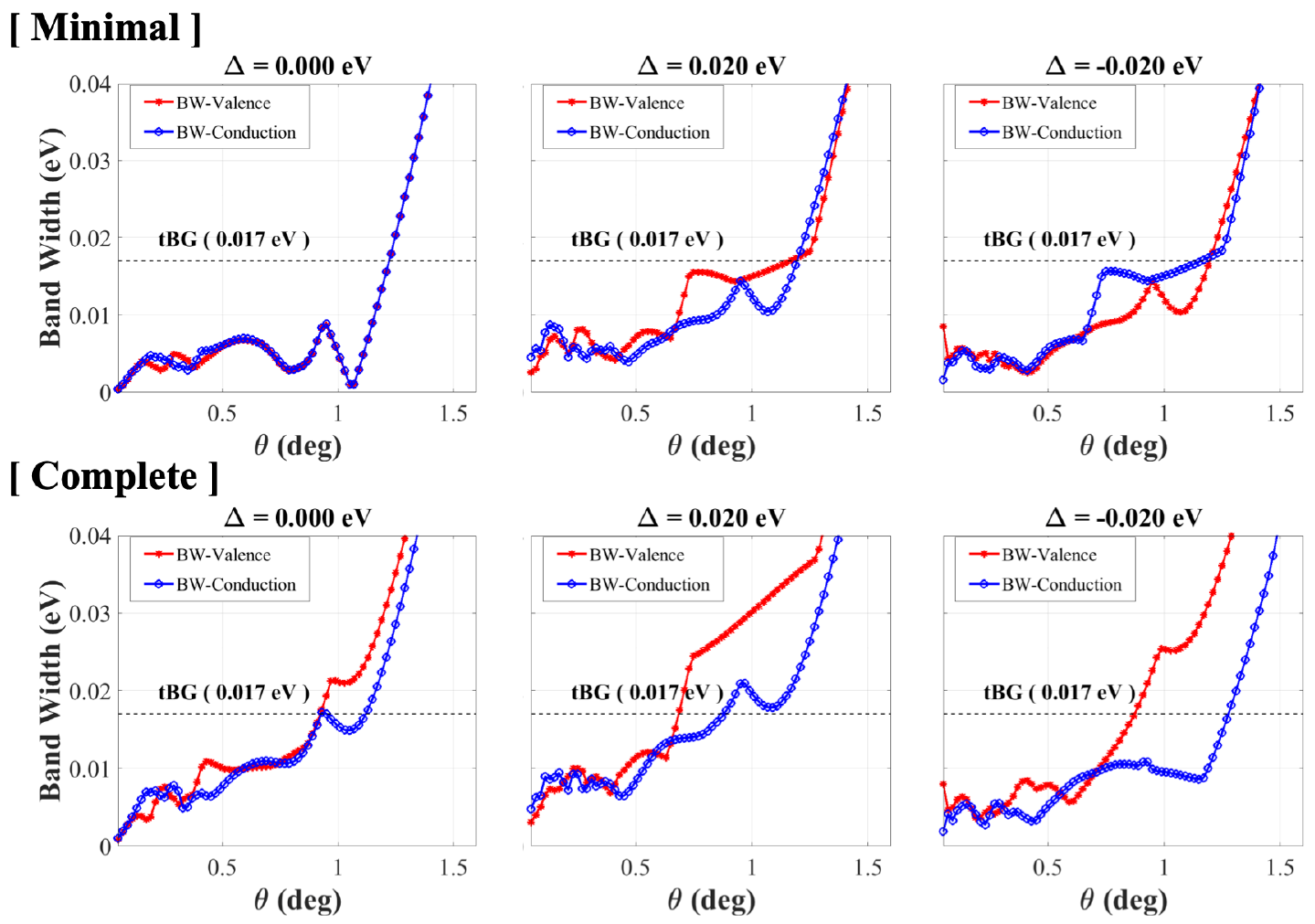} 
\caption{(color online) The variation of bandwidth as a function of twist angle for the valence and conduction bands within the minimal (top panel) and complete (bottom panel) models in twisted monolayer graphene on bilayer graphene (tMBG). We compare the bandwidth variations for applied potential $\Delta$ between the layers. Extremely small bandwidths are associated to the twist angle $\theta = 1.07^{\circ}$ in the minimal model for $\Delta = 0$ eV, which are a factor two smaller than the twisted bilayer graphene (tBG) with similar system parameters, where the dotted horizontal lines represent the bandwidth of tGB at the belly maximum. However, the bandwidths increase due to the remote hopping terms in the complete model.}
\label{Fig:BW-supple1}
\end{figure*}

\section*{A2. Bandwidth phase diagrams }
Identification of isolated low energy bands is important in order to achieve strongly correlated phases. 
In the presence of a primary gap near charge neutrality given by $\delta_p= CB_{\rm min} -VB_{\rm max}$, the isolation of the flat bands from higher energy bands can
be identified through a secondary gap.  For the $n^{th}$ valence band the secondary gap is $\delta_s(V^n) = V^n_{\rm min} -V^{n-1}_{\rm max}$ and similarly for the $n^{th}$ conduction band it is $\delta_s(C^n) = C^{n+1}_{\rm min} - C^n_{\rm max}$. The positive value of $\delta_s$ indicates  isolation and a negative value indicates overlap with the higher energy bands. 
It is understood from appendix~A1 that the bandwidths of valence/conduction bands in tMBG are tunable with the electric field $\Delta$ and depends on the sign of $\Delta$. In Fig.~\ref{Fig:BW_GAP-supple}, we show the valence/conduction bands bandwidth, secondary, and primary gaps phase diagrams for the parameter space of twist angle ($\theta$) and electric field ($\pm\Delta$). It is clear that (See Fig.~\ref{Fig:BW_GAP-supple} (a) and (b)), the bandwidth minima do not evolve linearly with  twist angle ($\theta$) nor does with electric field ($\pm\Delta$) in both minimal and complete models. 
The primary gap vanished in the complete model for most of the positive electric fields $\Delta$. 
In the main text we introduced a staggered potential $\Delta_S$ to open the gap of the monolayer graphene's linear bands of tMBG which has also an effect in the band isolation and bandwidth minimum. In Fig.~\ref{Fig:BW_GAP-supple} (c) and (d) for the complete model, we show respectively the effects of positive and negative staggered potentials. In the presence of a staggered potential ($\Delta_S$), the parameter space for the primary gap is increased, as well as the secondary gaps in the conduction bands for positive electric fields.
\begin{figure*}[!htb]
\includegraphics[width=17cm,angle=0]{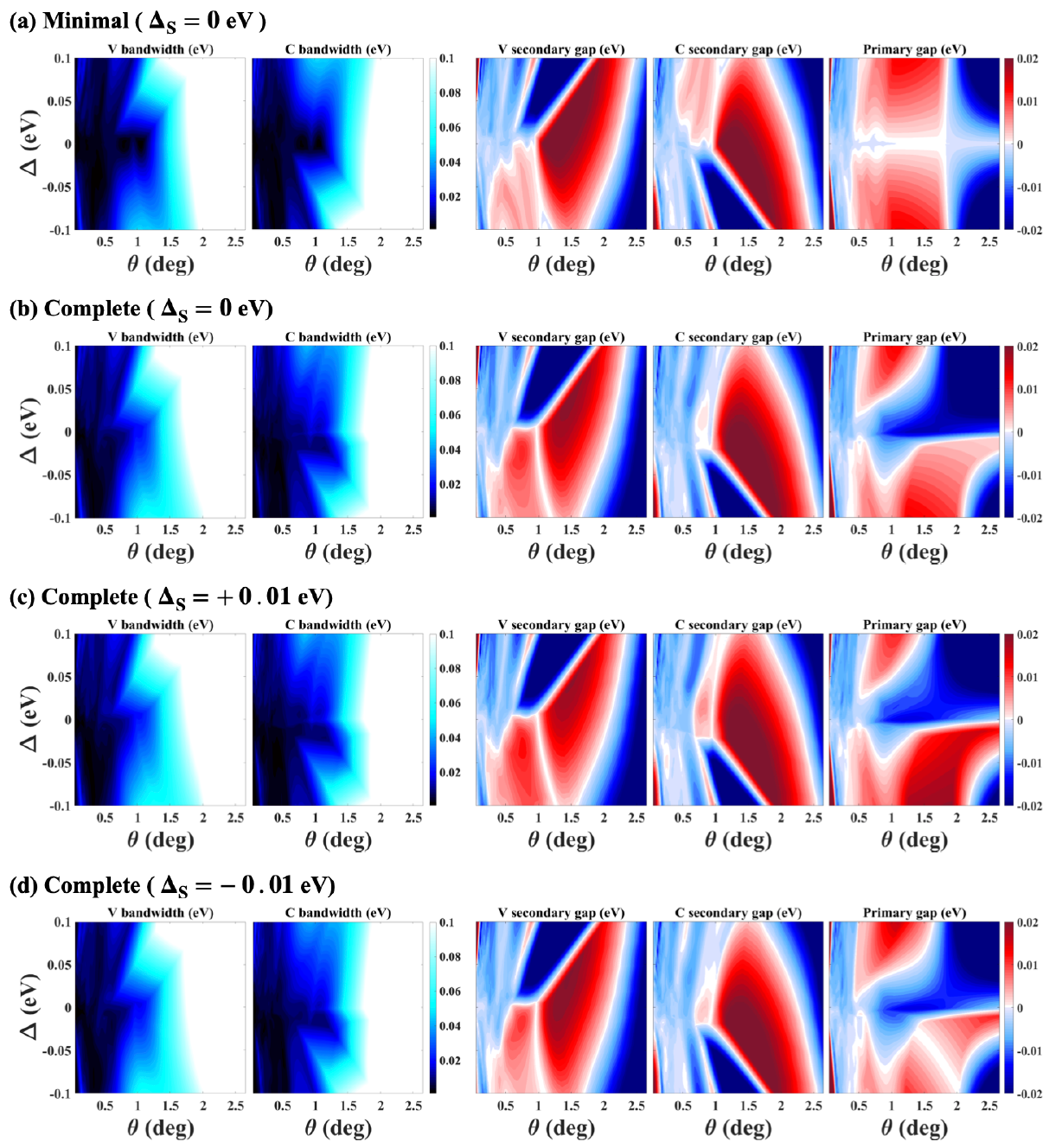} 
\caption{(color online) 
Colormap distribution of flat band bandwidth ($W$), the primary ($\delta_p$) and secondary ($\delta_s$) band gaps for the low-energy valence ($V$) and conduction ($C$) bands of tMBG in the parameter space of $\theta$ and $\Delta$. The (a) and (b) panels show the phase diagrams obtained using the minimal and complete models respectively for the staggered potential $\Delta_S = 0$~eV and the (c) and (d) panels show the results obtained using the complete model with the staggered potential $\Delta_S = \pm 0.01$ eV.}
\label{Fig:BW_GAP-supple}
\end{figure*}
%

\section*{A3. Promising twist angles and band structures}
The electronic structure for promising flat band system parameters are shown in Fig.~\ref{Fig:Bands-supple}
for select twist angles (0.51$^{\circ}$, 0.85$^{\circ}$, 1.21$^{\circ}$, 1.31$^{\circ}$,  and 1.41$^{\circ}$) for $\Delta$ values that  isolate either the conduction or the valence flat bands. The bandwidth and isolation of the low-energy bands of tMBG depend on the sign of electric field $\Delta$. The valley Chern number of the isolated bands are sensitive to the sign and magnitude of $\Delta$.

\begin{figure*}[!htb]
\includegraphics[width=14cm,angle=0]{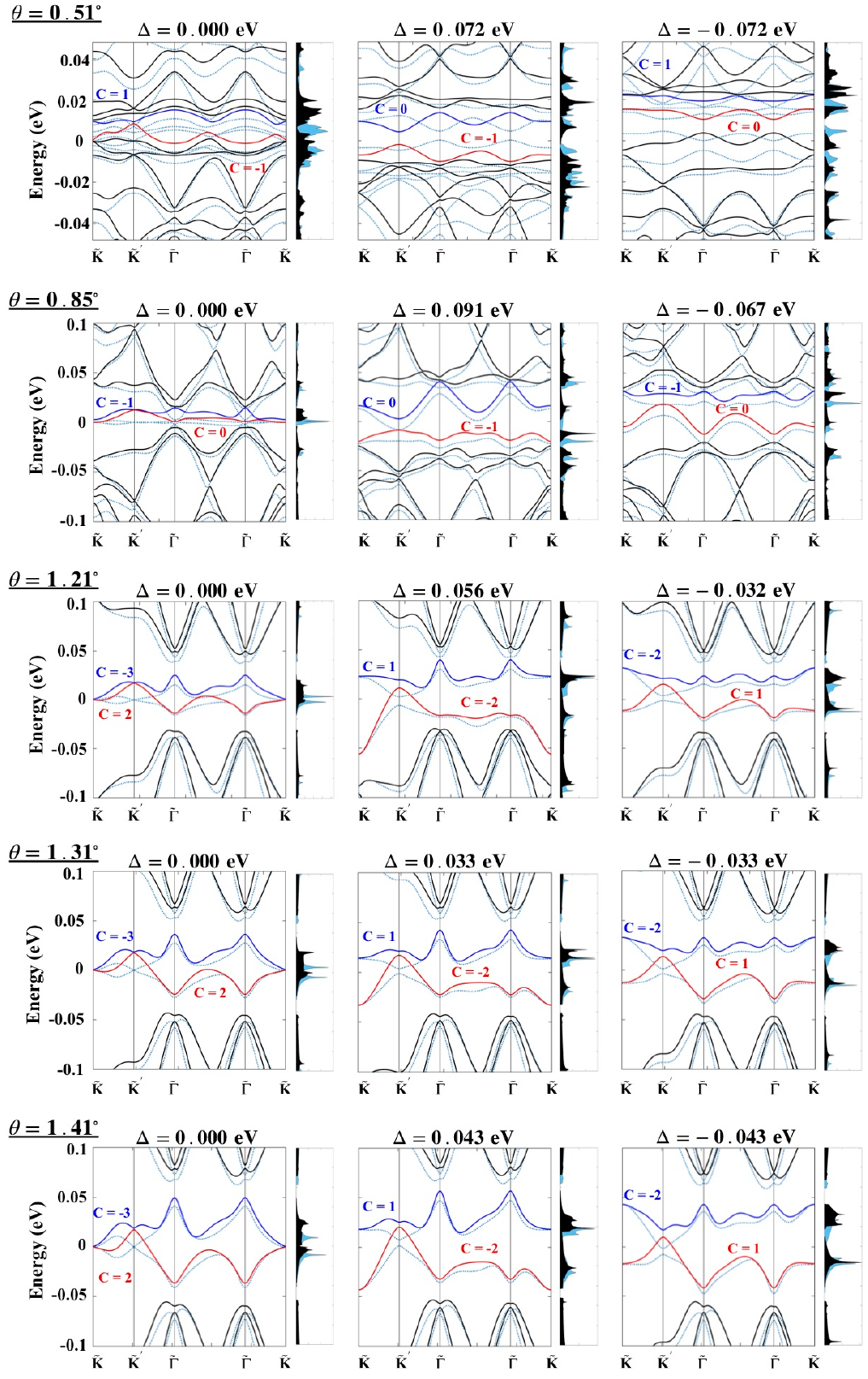} 
\caption{(color online) 
Band structure and density of states (DOS) of tMBG for select twist angles 0.51$^{\circ}$, 0.85$^{\circ}$, 1.21$^{\circ}$, 1.31$^{\circ}$,  and 1.41$^{\circ}$ and finite interlayer potential $\Delta$ that maximizes correlations for either the conduction or valence bands such that $U_{\rm eff}/W>1$. We compare the band structures and DOS for given twist angles obtained using minimal (light blue) and complete (black) models. In each band structure within the complete model, the low energy valence band is identified with red and the  conduction band with blue. The calculated Chern numbers within the complete model are shown together with the respective low energy bands using the same color. }
\label{Fig:Bands-supple}
\end{figure*}

\section*{A4. Effect of staggered potential ($\Delta_S$) on valley Chern numbers}
The electric field $\Delta$ is not sufficient to open the gap between the linear bands associated to the monolayer graphene in tMBG. In the main text, we defined a staggered potential ($\Delta_S$) to introduce a gap at the monolayer linear bands 
that enhanced the isolation and further reduced the bandwidth of low the energy bands of tMBG. 
From appendix~A3, it is known that the topology of the nontrivial bands can be modified with $\pm\Delta$.
Here we show that the staggered potential ($\Delta_S$) can also impact the valley Chern numbers of the low energy bands
for appropriate twist angle $\theta$ and interlayer potential $\Delta$
by comparing the calculations of the electronic structure for specific twist angles 1.21$^{\circ}$ and 1.41$^{\circ}$ 
for different sets of $\Delta$ and $\Delta_S$ values shown in Figs.~\ref{Fig:Bands-supple1}~\ref{Fig:BandsChern}.

\begin{figure*}[!htb]
\includegraphics[width=17.5cm,angle=0]{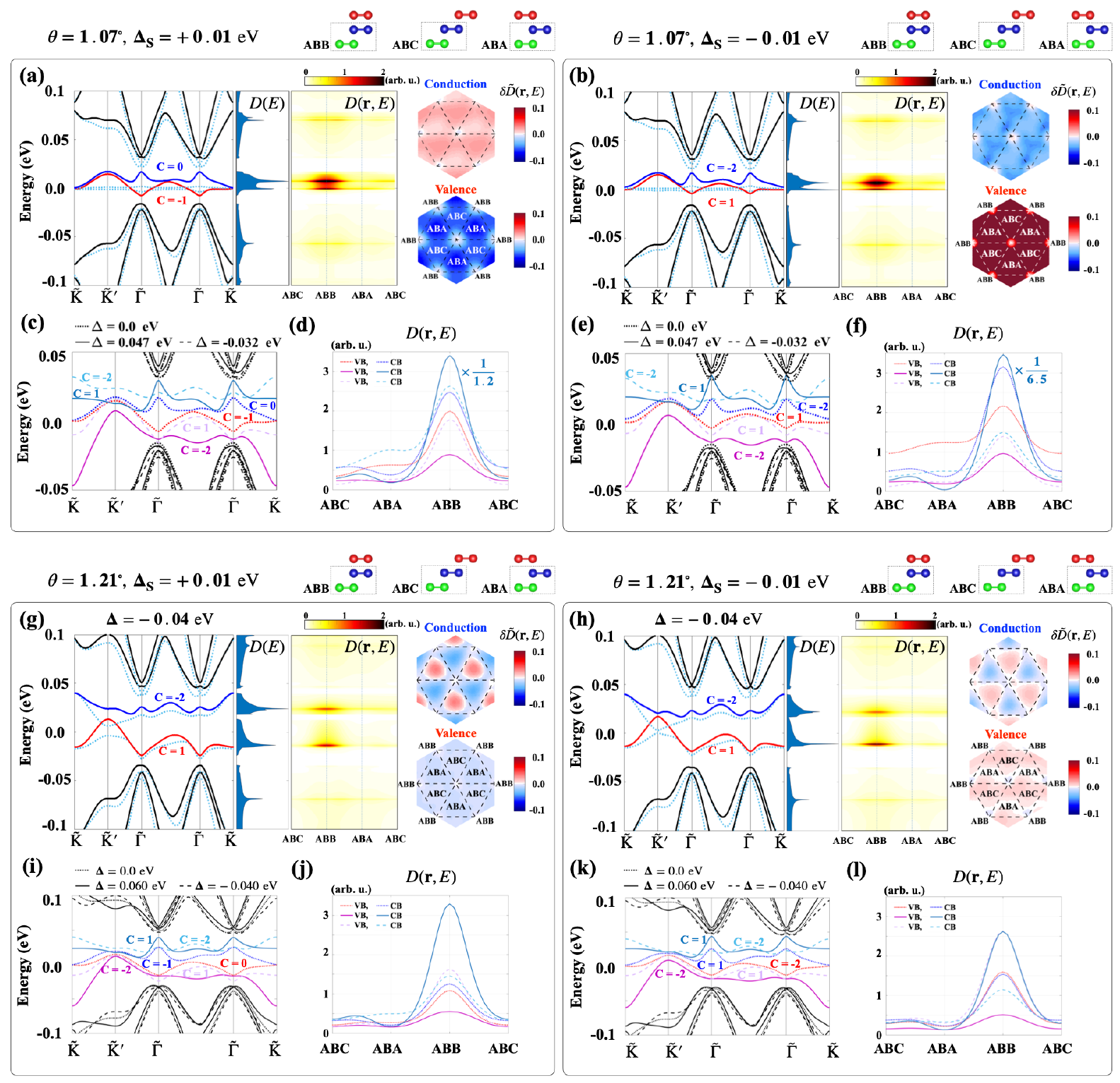} 
\caption{(color online) 
Band structures for opposite staggered potential signs $\Delta_S = \pm 0.01$~eV are shown 
on the left (a), (c), (d)  and right (b), (e), (f) panels for a fixed twist angle $\theta = 1.07^{\circ}$
corresponding to the magic angle of the minimal model for various interlayer potentials $\Delta$.
Similar plots can be found for (g), (i), (j) and (h), (k), (l) panels for twist angle $\theta = 1.21^{\circ}$.
Minimal and complete model bands are shown through dotted and solid lines respectively, 
and for the complete model the density of states $D(E)$, the local density of states $D({\mathbf r},E)$, and the difference 
$\delta \tilde{D}({\mathbf r},E) = \tilde{D}_{\Delta_S}({\mathbf r},E) - \tilde{D}_{0}({\mathbf r},E)$ due to the 
staggering potential $\Delta_S$ where
$\tilde{D}({\mathbf r}, E) = D({\mathbf r},E) / {\rm max}(| D({\mathbf r},E)| )$ are the normalized LDOS evaluated at 
the van Hove singularity energies $E = E_{\rm VHS}$ of the nearly flat bands.
The electronic wave functions are found to concentrate maximally for conduction bands at ABB stacking regions.
}
\label{Fig:Bands2}
\end{figure*}

\section*{A5. Sub-lattice and twisted layer resolved local density of states at twist angle $\theta = 1.07^{\circ}$}
The LDOS at the twist angle $\theta =1.07^{\circ}$ are mainly located at the ABB stacking in tMBG regardless of the value for $\Delta_S$. 
In Fig.~\ref{Fig:LDOS-supple}, we further show by projecting the LDOS onto the sub-lattices 
that the charge localization at the ABB stacking concentrate mainly at the low energy 
A$_b$ and B$_m$ sites of the bilayer and the vertically contiguous B$_t$ of the monolayer.
The low energy sub-lattice B$_m$ of the bilayer has the dominant contribution over all the low energy sub-lattices.
In the presence of an applied electric field, the localization completely polarized to the sub-lattices B$_t$ and B$_m$,
and further inclusion of a staggered potential $\Delta_S$ influences the localization where a negative 
$\Delta_S$ increases the LDOS on sub-lattices B$_t$ and B$_m$ by almost an order of magnitude.
We expect the preferential occupation of the low energy sites by the flat bands for other twist angles while 
changes introduced by $\Delta_S$ may behave differently. 
This type of electron localization behaviors could in principle be observed through scanning probe measurements.

\begin{figure*}[!htb]
\includegraphics[width=17.5cm,angle=0]{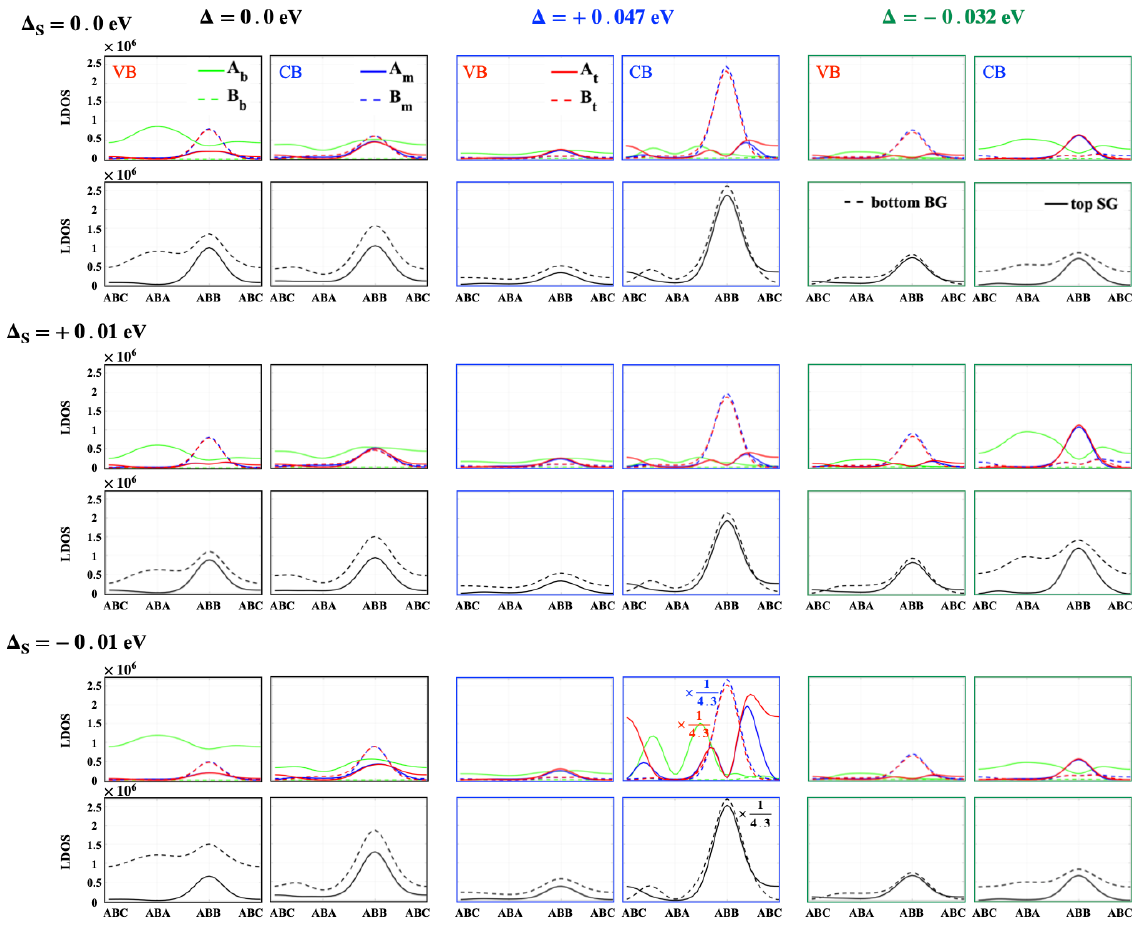} 
\caption{(color online) 
Sub-lattice and layer projected LDOS at valence/conduction VHS for twist angle $\theta = 1.07^{\circ}$ 
for different combination of $\Delta=0.0,\, 0.047,\, -0.032$~eV and $\Delta_S=0.0,\,0.01,\,-0.01$~eV. 
The low energy sub-lattices (B$_m$ of the bilayer and B$_t$ of monolayer) are found to 
concentrate most of the flat bands charge density at the van Hove singularity. 
}
\label{Fig:LDOS-supple}
\end{figure*}

\begin{figure*}[!htb]
\includegraphics[width=17cm,angle=0]{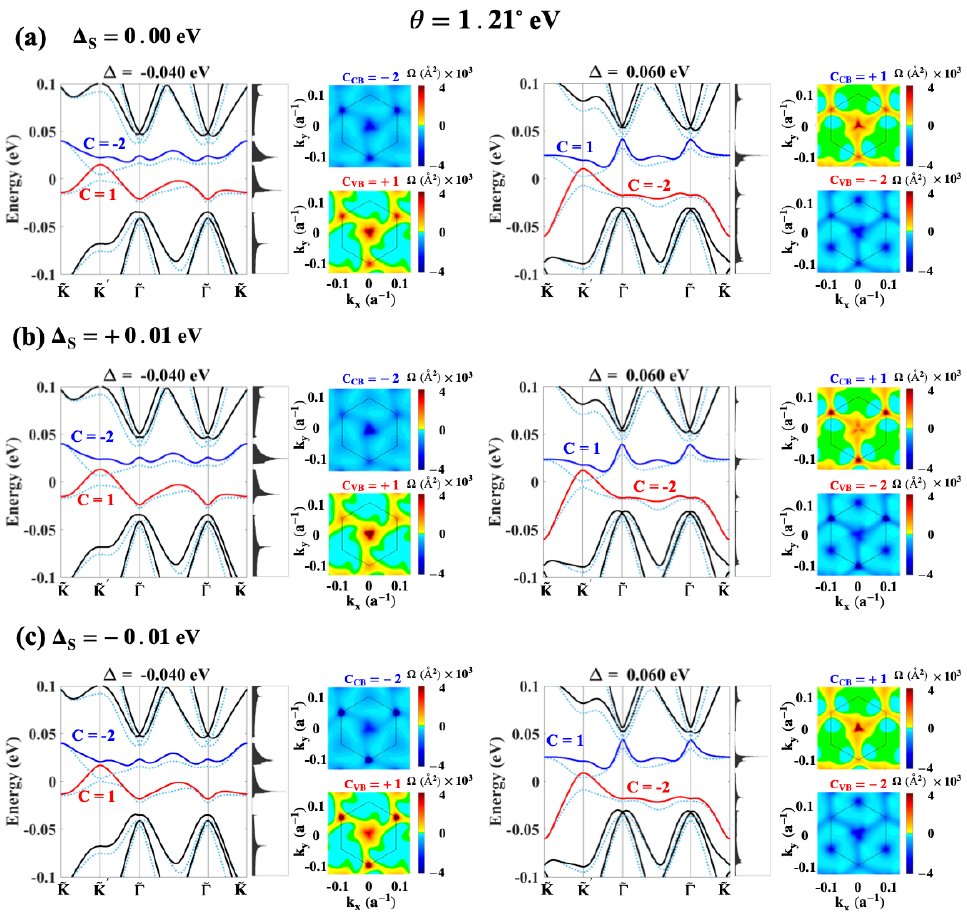} 
\caption{(color online) 
Band structures and berry curvatures of tMBG for the twist angle $\theta = 1.21^{\circ}$  with applied electric field $\Delta = \pm 0.02$~eV.  We compare the minimal (light blue) and complete (black) models for electric fields $\Delta = \pm 0.02$~eV with (a) the staggered potential set to $\Delta_S = 0$, (b) $\Delta_S = 0.01$~eV, and (c) $\Delta_S = -0.01$~eV 
that do not show change the valley Chern numbers for cases explored. 
For each case of $\Delta$ and $\Delta_S$, we show the berry curvatures plot. 
In each surface plot, the first moire BZ is indicated by the black hexagonal lines. 
The berry curvature hotspots are observed at ${\rm \tilde{K}^{\prime}}$ and ${\rm \tilde{\Gamma}}$ within the MBZ.}
\label{Fig:Bands-supple1}
\end{figure*}

\begin{figure*}[!htb]
\includegraphics[width=17cm,angle=0]{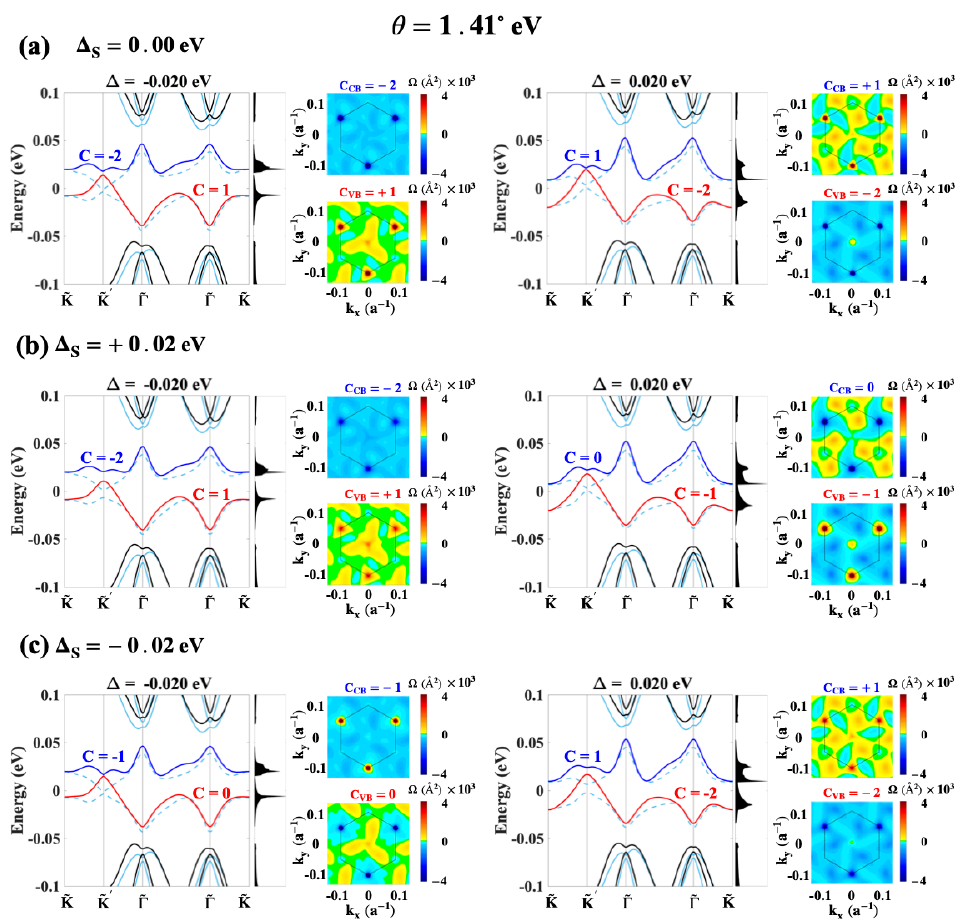} 
\caption{(color online) 
Band structures and berry curvatures of tMBG for the twist angle $\theta = 1.41^{\circ}$  with applied electric field induced potentials of $\Delta = \pm 0.02$~eV. 
We compare the minimal (light blue) and complete (black) models for electric field induced potentials
$\Delta = \pm 0.02$~eV for (a) staggered potential set to $\Delta_S = 0$ and for non-zero staggered potential (b) $\Delta_S = 0.01$~eV, and (c) $\Delta_S = -0.01$~eV. The low energy bands show variations in the valley Chern numbers compared to the case of $\Delta_S = 0$. For each case of $\Delta$ and $\Delta_S$, we show the berry curvatures plot. In each surface plot, the first moire BZ is indicated by the black hexagonal lines. The berry curvature hotspots are observed at ${\bm \tilde{K}^{\prime}}$ and this is due to the small gap opening of the linear bands of monolayer graphene in tMBG.
}
\label{Fig:BandsChern}
\end{figure*}

\end{appendix}

\end{document}